\documentclass[a4paper,12pt]{scrartcl}
\usepackage[utf8x]{inputenc}			
\usepackage[T1]{fontenc}				
\usepackage[english]{babel}				
\usepackage{color}						
\usepackage{amsmath}					
\usepackage{amstext}					
\usepackage{amssymb}
\usepackage{amsfonts}					
\usepackage{amsthm}						
\usepackage{mathrsfs}					
\usepackage{dsfont}
\usepackage{stackrel}					
\usepackage[round]{natbib} 
\usepackage{csquotes}
\usepackage{microtype}
\usepackage{fourier}
\usepackage{hyperref}
\usepackage{mathtools}
\usepackage{graphicx}
\usepackage{subcaption}
\usepackage[tikz]{mdframed}
\usetikzlibrary{positioning}

\makeatletter
\g@addto@macro{\thm@space@setup}{\thm@headpunct{}}
\makeatother



\newtheorem{conclusion}[]{Conclusion}
\theoremstyle{remark}

\numberwithin{equation}{section}
\numberwithin{definition}{section}
\allowdisplaybreaks[4]

\newmdenv[innerlinewidth=0.5pt, roundcorner=6pt,linecolor=black,backgroundcolor=yellow!20,innerleftmargin=6pt,
innerrightmargin=6pt,innertopmargin=-0pt,innerbottommargin=6pt]{mybox}

\makeatletter
\let\orig@mybox=\mybox
\def\mybox#1{
  \orig@mybox[frametitle={#1}]
}
\makeatother

\newcounter{defcounter}
\newenvironment{boxequation}{%
\addtocounter{equation}{-1}
\refstepcounter{defcounter}

\begin{equation}}
{\end{equation}}

\title{Understanding evolutionary and ecological dynamics using a continuum limit} 
\subtitle{A methods review for stochastic differential equations}
\author{Peter Czuppon$^{1,2,3}$, Arne Traulsen$^3$}
\date{}

\begin{document}

\maketitle

\vspace{-30pt}

\begin{center} \small
\noindent{} $^1$ \textit{Centre Interdisciplinaire de Recherche en Biologie, CNRS,\\ Coll\`ege de France, PSL Research University, Paris (France)}\vspace{5pt}

$^2$ \textit{Institute of Ecology and Environmental Sciences Paris, UPEC, CNRS, \\ IRD, INRA, Sorbonne Universit\'{e}, Paris (France)}\vspace{5pt}

$^3$ \textit{Department of Evolutionary Theory,\\ Max Planck Institute for Evolutionary Biology, Pl\"on (Germany)}

\end{center}

\vspace{-20pt}
\subsection*{Disclaimer}
\vspace{-7pt}
This manuscript contains nothing new, but synthesizes known results:
For the theoretical population geneticist with a probabilistic background, we provide a summary of some key results on stochastic differential equations. 
For the evolutionary game theorist, we give a new perspective on the derivations of results obtained when using discrete birth-death processes. 
For the theoretical biologist familiar with deterministic modeling, we outline how to derive and work with stochastic versions of classical ecological and evolutionary processes.

\vspace{-12pt}
\tableofcontents

\section{Introduction}
The huge computational power available today allows more and more theoreticians to develop individual-based models of high complexity to explore dynamical processes in ecology and evolution. Here, we aim to make a link between these individual-based descriptions and continuous models like (stochastic) differential equations that remain amenable to analysis. We review these techniques and apply them to some frequently used models in ecology and evolution. 

In ecology, probably the most common description of population dynamics is the logistic growth equation \citep{verhulst:Quetelet:1838}. Its attractiveness draws from its simplicity. It has a globally attractive fixed point (when started with any non-zero population size), the carrying capacity of the population. However, this simplicity comes at the cost that biological observations like population size fluctuations or even extinction events are not captured by this deterministic model. To account for these stochastic effects one needs to change to a stochastic differential equation which can be derived from individual-based reactions \citep{champagnat:TPB:2006}. We outline this procedure along similar lines as reviewed in~\citet{black:TREE:2012} but with a larger emphasis on the technical details.  

In evolutionary game theory, the Moran process has become a popular model for stochastic dynamics in finite populations \citep{nowak:Nature:2004}. It is a model describing the dynamics of different alleles in a population of fixed size and overlapping generations.
As this is a birth-death process, quantities like fixation probabilities, times, and the stationary distribution can be calculated based on recursions \citep{goel:book:1974,karlin:book:1975,traulsen:bookchapter:2009a,allen:book:2011}. A continuum approximation for quantities that are known exactly may thus make limited sense at first sight. 
Another important process in population genetics is the Wright-Fisher process -- a model for allele frequency dynamics in a population of fixed size and non-overlapping generations \citep{wright:Genetics:1931}. It is more popular in population genetics but is also used in evolutionary game theory \citep[e.g.][]{imhof:JMB:2006,traulsen:PRE:2006c,taylor:PRSB:2012,wakano:JTB:2014}. 
The Wright-Fisher process is mathematically more challenging to analyze exactly than the Moran process. 
Therefore, continuum approximations resulting in stochastic differential equations are used to compute typical quantities of interest such as the probability of fixation of a certain genotype or the mean time until this fixation event occurs \citep{crow:book:1970,ewens:book:2004}. Additional to the derivation of the continuum limit, we also present how to compute these quantities. 

Even though similar in the questions they try to answer, evolutionary game theory and population genetics are developing in parallel, sometimes with little interaction between them. 
As this is partly arising from the different methods applied, here
we aim to provide an introduction to the continuum limit for those less comfortable with these methods and hesitant to go into the extensive, more mathematical, literature. 

Since our goal is to illustrate how to apply a continuum limit to individual-based descriptions of a biological process, the calculations and derivations below may remain vague where more mathematical theory is necessary. 
For a mathematically rigorous presentation of this topic we refer to the excellent lecture notes by \citet{etheridge:LN:2012} or the book by \citet{ewens:book:2004}. 
A more application-oriented treatment of stochastic processes in biology can be found in the books by \citet{otto:book:2007} and \citet{allen:book:2011}.

\section{Evolutionary and ecological proto-type processes}\label{sec:models}
We outline the derivation of continuum limits by application to exemplary processes from evolution and ecology; the Wright-Fisher and Moran process, and the logistic equation. By showing the explicit derivation in these examples, we provide the necessary tool set to derive continuum limits of more complex processes motivated by individual-based models. In this section we define the models by their microscopic descriptions, that is, we describe the model dynamics as viewed from an individual's perspective.

\subsection{Wright-Fisher and Moran process}\label{sec:wf}
The two most popular processes to model (stochastic) evolutionary dynamics are the Wright-Fisher and the Moran process. While in the Wright-Fisher process generations are non-overlapping and time is measured in discrete steps, generations in the Moran model are overlapping and measured in either discrete or continuous time. Originally, both processes describe the stochastic variation of allele frequencies due to finite population size effects referred to as \textit{genetic drift}. 

\subsubsection{Wright-Fisher model}
One of the oldest population genetics model is the finite size Wright-Fisher process \citep{fisher:book:1930,wright:Genetics:1931}. Given a population of constant size, it describes the change in frequencies of alleles in non-overlapping generations over time, measured in (discrete) generations.

Classically, one considers a population of $N$ individuals where each individual is of type $A$ or $B$. The population is considered to be in its ecological equilibrium. The population size $N$ is therefore constant over time. 
One interpretation of the dynamics is that every generation each individual chooses, independently of all other individuals, an ancestor from the previous generation and inherits its type. Under selection, the likelihood of drawing type $A$ individuals increases (or decreases) which introduces a sampling bias. The probability for an offspring to have a parent of type $A$, conditional on $k$ individuals being of type $A$ in the parental generation, is
\begin{equation}\label{eq:selection_probability}
	p_k = \frac{(1+s)k}{(1+s)k + N-k},
\end{equation}
where $s\in\mathbb{R}_{\geq 0}$ is the selective advantage of type $A$.
The number of type $A$ individuals in the next generation is then given by a binomial distribution with sample size $N$ and success probability $p_k$. Denoting the number of type $A$ individuals in generation $n$ by $X_n$ we have
\begin{equation}\label{eq:wf_prob}
	\mathbb{P}(X_{n+1} = j |X_n=k) = \binom{N}{j} p_k^j (1-p_k)^{N-j},\quad 0\leq k\leq N.
\end{equation}

Unfortunately the Wright-Fisher model, even though very illustrative, is difficult to study analytically. 
Through the developments in stochastic modeling in the last century, a lot of this new theory could be adopted to overcome this problem \citep[e.g.][]{kimura:book:1983,ewens:book:2004}. 

\subsubsection{Moran model}
Another way to resolve the difficulties associated with the Wright-Fisher process is provided by the Moran process \citep{moran:MPCPS:1958}. The setup is the same as for the Wright-Fisher process (constant population size $N$ with two types or -- in population genetics -- alleles $A$ and $B$) with one exception: time is not measured in generations but each change in the population configuration affects only one individual, the one that dies and gets replaced by an offspring of another randomly selected individual. Therefore, generations are overlapping and time can be measured in discrete steps or continuously. 

\subsubsection*{Discrete time}
The Moran process in discrete time progresses as follows. Every time step, one individual is randomly chosen to reproduce and the offspring replaces a randomly chosen individual among the remaining $N-1$ individuals (sometimes the replacement mechanism is not restricted to the remaining individuals but also includes the parent). Therefore, in a population with $k$ type-$A$ individuals, the probability that one of these replaces a type-$B$ individual is given by 
\begin{equation} 
	T^{k+} = p_k \frac{N-k}{N-1},\quad \text{for } 0\leq k\leq N\quad \Big(\text{sometimes } T^{k+} = p_k \frac{N-k}{N} \Big)
\end{equation}
with $p_k$ as defined in Eq.~\eqref{eq:selection_probability}. Analogously, the probability for the number of type-$A$ individuals to decrease from $k$ to $k-1$ reads
\begin{equation}
	T^{k-} = (1-p_k) \frac{k}{N-1},\quad \text{for } 0\leq k \leq N .
\end{equation}

We have implemented selection on the reproduction step, but it could also affect the replacement step. In a non-spatial setting, as considered here, this leads to the same transition probabilities. 
However, the Moran model can also be studied on a graph which aims to model spatial structure. In that case, the order of reproduction and replacement, and which of these steps is affected by selection, matters and can potentially give rise to different evolutionary dynamics \citep{lieberman:Nature:2005,kaveh:JRSOS:2015}. 
We note further that without selection ($s=0$) we have $p_k=k/N$, i.e. the increase and decrease probabilities are equal for any choice of $k$. Dynamics with this property are called \textit{neutral}.

\subsubsection*{Continuous time} 
The same dynamics (albeit on a different time-scale) are obtained by assuming that each pair of individuals is associated with a random exponentially distributed time (also described as exponential clocks). The next pair to update their types is determined by the smallest random time (or the clock that rings first). At these updating times, one of the two individuals is chosen to reproduce, the offspring replacing the other individual of the pair. There is no standard choice when it comes to choosing the rate of these exponential times. \\

Both formulations of the Moran process are Markov chains, either in discrete or continuous time, with the special property of having jumps of $\pm 1$ only. These processes are called \textit{birth-death processes}. The theory of these is well developed, see for example the books of \citet{karlin:book:1975,karlin:book:1981}, \citet{gardiner:book:2004}, or \citet{allen:book:2011}, so that the dynamics of Moran processes are often amenable to analysis (typically by solutions of recursion equations).

\begin{conclusion}
The difference between the Wright-Fisher model and the Moran model is the progression of populations in time. In the Wright-Fisher process, generations are non-overlapping, i.e.\ all individuals update their type at the same time. Therefore the distribution of types in the offspring generation is binomial. In contrast, generations are overlapping in the Moran model and the dynamics are described by a birth-death process since only one individual is updated at a time. 
\end{conclusion}

\subsection{Logistic growth}
In ecology one is typically interested in population sizes or densities rather than allele frequencies. The simplest population growth model is that of exponential growth. 
Obviously, a population cannot grow exponentially forever. Its growth will be limited at some point, for example due to spatial constraints or resource depletion. This form of density regulation suffices to stabilize a population around its carrying capacity, the positive population size at which in the deterministic process the growth rate equals zero. 

Here, we give the mechanistic basis that could potentially describe such a process. We denote a single individual of the population by $Y$. The birth- and death-processes can then be written as
\begin{equation}\label{eq:logistic_bd}
	\begin{aligned}
		Y &\stackrel{\beta}{\longrightarrow} Y + Y,&\quad &\text{birth};\\
		Y &\stackrel{\delta}{\longrightarrow} \varnothing,&\quad &\text{death}.	
	\end{aligned}
\end{equation}
The parameters $\beta$ and $\delta$ correspond to the rates at which the two reactions happen, i.e. each reaction corresponds to an exponential clock with rate either $\beta$ or $\delta$. For $\beta>\delta$, the population grows to infinity (exponential growth), whereas for $\beta<\delta$, it goes extinct. 

Population regulation is achieved through a non-linear term that is interpreted as an interaction between two individuals, e.g. competition for space. The corresponding microscopic process is given by
\begin{equation}\label{eq:logistic_comp}
	Y + Y \stackrel{\gamma/K}{\longrightarrow} Y, \qquad \text{(competition)}.
\end{equation}
The parameter $\gamma$ is referred to as the intra-specific competition coefficient and $K$ is a measure of the number of individuals at carrying capacity. 
The division by $K$ in the competition rate is accounting for the probability of interaction of two individuals in a well-mixed population where space is measured by the parameter $K$ so that $Y/K$ becomes a density (or rate of encountering an individual when randomly moving in space). For a more detailed derivation of these type of interaction rates we refer to \citet{anderson:book:2015}.

The logistic process is, like the Moran process, a birth-death process. 
We will see in the next section, that in the infinite population size limit (we let $K$ tend to infinity) the mechanistic description above yields the logistic equation
\begin{equation}
	\frac{dy}{dt} = r y \left(1-\frac{y}{c}\right),
\end{equation}
where $r=\beta-\delta$ is the per-capita growth-rate, $c=(\beta-\delta)/\gamma$ is the rescaled carrying capacity, and $y= Y/K$ is the density of the population.

\section{Infinite population size limit}\label{sec:infinite_limit}
The microscopic descriptions can be implemented by a stochastic simulation algorithm. Yet, the theoretical analysis of finite size populations can be challenging. A common technique to overcome this challenge is to consider a continuum approximation, i.e. studying the limiting model for $N$ (or $K$) to infinity. 
The limit is a (stochastic) differential equation of the form
\begin{equation}\label{eq:sde}
	dx_t = \mu(x_t)dt + \sigma(x_t)dW_t,
\end{equation}
where $(W_t)_{t\geq 0}$ is a standard Brownian motion (see Box~1). This equation describes the population dynamics, i.e. the macroscopic evolution of a certain model. For a general introduction to stochastic differential equations, see for example the books by \citet{karlin:book:1981} and \citet{gard:book:1988}.

\begin{figure*}[t!]
	\begin{mybox}{Box 1: Brownian motion}
	\small An intuitive way to think about a Brownian motion is its discrete analog (both in time and space), the symmetric random walk. Every time step the process increases by $1$ with probability $1/2$ and decreases by $1$ with probability $1/2$ (see Fig. (a) in this box). Rescaling time and space, so that both approach continuous quantities, the random walk converges to a standard Brownian motion, a process that is at every time point $t$ normally distributed with mean $0$ and variance $t$ \citep[e.g.][Theorem~14.9]{kallenberg:book:2002}. Formally, the Brownian motion $(W_t)_{t\geq 0}$ is an $\mathbb{R}$-valued stochastic process where for $0=t_0<t_1<t_2<...<t_n$ the increments $(W_{t_i}-W_{t_{i-1}})_{i\in\{1,...,n\}}$ are independent of each other and distributed according to $\mathcal{N}(0,t_i-t_{i-1})$. Here, $\mathcal{N}(\mu,\sigma^2)$ is a normal distribution with mean $\mu$ and variance $\sigma^2$. The standard Brownian motion is distributed as $W_t\sim\mathcal{N}(0,t)$.
	
	\begin{subfigure}{.5\textwidth}
  		\centering
  		\includegraphics[width=\linewidth]{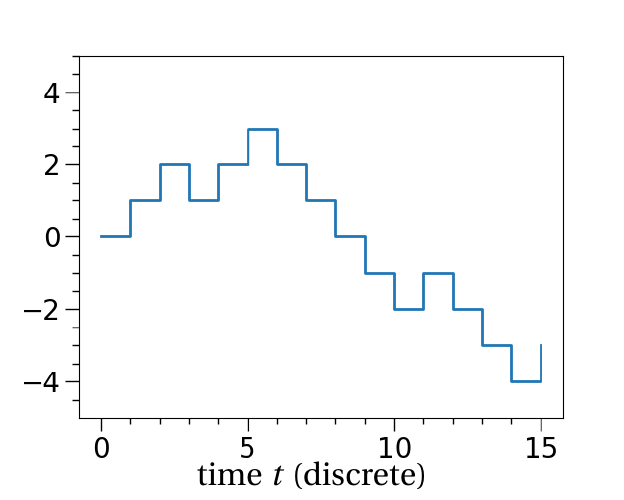}
  		\caption{random walk}
	\end{subfigure}%
	\begin{subfigure}{.5\textwidth}
 		 \centering
 		 \includegraphics[width=\linewidth]{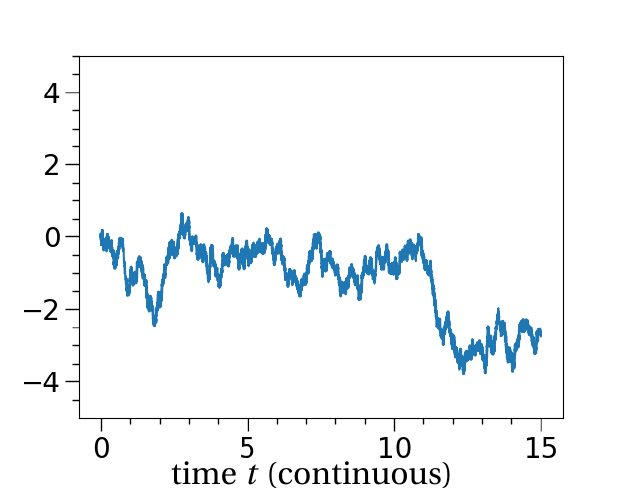}
  		\caption{Brownian motion}
	\end{subfigure}	
		

	\end{mybox}
\end{figure*}

\setcounter{figure}{0}

The term $\mu(x_t)$ is called the infinitesimal mean, i.e. the expected change of the stochastic process $(x_t)_{t\geq 0}$ in a very short (infinitesimal) time interval. It represents the deterministic dynamics of the process. The term $\sigma^2(x_t)$ is called the infinitesimal variance, i.e. the expected variation of the continuum limit in very small time steps. It quantifies the random fluctuations. In the case where $\sigma$ is zero, the limit is deterministic and Eq.~\eqref{eq:sde} reduces to an ordinary differential equation. 

One can formally show that the terms $\mu(x)$ and $\sigma^2(x)$ indeed correspond to the changes of the mean and the variance in infinitesimally small time steps if they are derived formally (see Appendix~\ref{app:sde}). This allows us to compute them by the following identities:
\begin{equation}\label{eq:inf_mean_var}
	\begin{aligned}
		 \mu(x) &= \lim_{\Delta t\to 0} \frac{1}{\Delta t} \mathbb{E}[(x_{\Delta t}-x_0)|x_0=x] \, ,\\
		 \sigma^2(x) &= \lim_{\Delta t\to 0} \frac{1}{\Delta t} \mathbb{V}[(x_{\Delta t}-x_0)|x_0=x] \, ,
	\end{aligned}
\end{equation}
where $\mathbb{E}$ and $\mathbb{V}$ denote the expectation and variance of 
the process $x_t$.

A solution of a stochastic differential equation of the form in Eq.~\eqref{eq:sde} is called a \textit{diffusion}. Another common representation of Eq.~\eqref{eq:sde} is the following integral equation	
\begin{equation}
	x_t = \int_0^t \mu(x_s)ds + \int_0^t \sigma(x_s) dW_s,
\end{equation}
where the stochastic integral is interpreted in the sense of It\^{o}. For a discussion of the different choices of stochastic integrals and their consequences in terms of modeling, see for example \citet{turelli:TPB:1977}.


We now present how to derive Eq.~\eqref{eq:sde} for the three introduced models. The strategy is rather simple: Compute the infinitesimal mean and variance as given in Eq.~\eqref{eq:inf_mean_var} for the individual-based model.
%
\subsection{Discrete-time derivation: Wright-Fisher model}\label{sec:wf_derivation}
For the reason of illustration we assume a Wright-Fisher model as outlined in Section~\ref{sec:wf} without selection, $s=0$. We need to compute the expectations in Eq.~\eqref{eq:inf_mean_var} using the probability distribution given in Eq.~\eqref{eq:wf_prob} (with $s=0$). Let us ignore the time step $\Delta t$ for the moment and simply compute the change in expectation and variance from one generation to the other. Writing $X_t$ for the number of individuals of type $A$ and setting $\Delta X_t = X_{t} - X_{t-1}$, we find
\begin{equation} 
	\begin{aligned}
		\mathbb{E}[\Delta X_t|X_{t-1}=k] &= \mathbb{E}[X_{t}|X_{t-1}=k] - k = N\frac{k}{N} - k = 0,
	\end{aligned} 
\end{equation}
where we used that the number of individuals of a certain type in the next generation is binomially distributed. 
Analogously, the infinitesimal variance is
\begin{equation}\label{eq:wf_inf_var}
	\begin{aligned}
		\mathbb{E}[(\Delta X_t)^2|X_{t-1}=k] &= \mathbb{E}[X_{t}^2|X_{t-1}=k] -2k\mathbb{E}[X_{t}|X_{t-1}=k] + k^2\\
		&= \mathbb{V}[X_{t}|X_{t-1}=k] +\mathbb{E}[X_{t}|X_{t-1}=k]^2 -2k^2 + k^2\\
		&= N \frac{k}{N} \left(1-\frac{k}{N}\right).
	\end{aligned}
\end{equation}
It remains to account for the transition from discrete to continuous time. In this case, the natural choice is $\Delta t=N^{-1}$. This can be seen by examining the infinitesimal variance. To obtain a limit different from zero or infinity in Eq.~\eqref{eq:wf_inf_var} for $N$ to infinity, we need to divide that equation by $N$. Comparing Eq.~\eqref{eq:wf_inf_var} to the corresponding line in Eq.~\eqref{eq:inf_mean_var}, we set $\Delta t=N^{-1}$. Replacing $k/N$ by $x$ and taking the limit $N\to\infty$, we find the neutral Wright-Fisher diffusion for the allele frequency dynamics
\begin{equation}\label{eq:wf-diff}
	dx_t =  \sqrt{x_t(1-x_t)} dW_t.
\end{equation}
This equation is called Wright-Fisher diffusion and describes the evolution of a neutral allele due to genetic drift. In other words, the allele frequency behaves like a random walk in continuous time and space. A similar derivation as above can be done by including selection and mutation. The more lengthy computation is relegated to Section~\ref{app:wf_selection} in the Appendix.

\begin{conclusion}
To derive a continuum limit of a finite population size model in discrete time, one computes the infinitesimal mean and variance as given in Eq.~\eqref{eq:inf_mean_var} and rescales time so that the two quantities converge in a meaningful way, i.e. do not tend to infinity.
\end{conclusion}

\subsection{Continuous-time derivation: General case}
In principle, the same methodology as above is applicable for the derivation of the continuum limit of a process measured in continuous time. However, in view of our subsequent analysis of the limit Eq.~\eqref{eq:sde}, we will introduce a new tool, the \textit{infinitesimal generator}.

The change in infinitesimal time of any continuous-time Markov process $(x_t)_{t\geq 0}$ can be described by the infinitesimal generator, denoted $\mathscr{G}$. 
Intuitively, one can think of it as the derivative of the expectation (of an arbitrary function) of a stochastic process. 
Formally, it is defined by
\begin{equation}\label{eq:inf_gen_def}
	(\mathscr{G}f)(x) = \lim_{\Delta t \to 0} \left( \frac{\mathbb{E}[f(x_{\Delta t})|x_0 = x] - f(x)}{\Delta t} \right),
\end{equation}
where $\mathbb{E}[f(x_{\Delta t})|x_0 = x]$ denotes the conditional expectation of the stochastic process $f(x_t)$ at time $\Delta t$ given the initial value $x_0=x$. Here, $f$ is an arbitrary function so that the limit is well-defined. For example, applying $\mathscr{G}$ to $f(x)=x$ describes the dynamics of the mean of~$x_t$, and for $f(x)=x^2$ we obtain the dynamics of the second moment of $x_t$. From the first two moments, we can recover the variance, so that from Eq.~\eqref{eq:inf_gen_def} one can derive the infinitesimal mean and variance. 

The infinitesimal generator is useful in our context since it can be related to a diffusion process. More precisely, the infinitesimal generator associated to the stochastic diffusion
\begin{equation}
	dx_t = \mu(x_t)dt + \sigma(x_t)dW_t 
\end{equation}
is given by (see Appendix~\ref{app:sde} for a derivation by the It\^{o} formula)
\begin{equation}\label{eq:sde_generator} 
	(\mathscr{G}f)(x) = \mu(x) f'(x) + \frac{1}{2}\sigma^2(x) f''(x). 
\end{equation}

Our strategy is to find a limit of the infinitesimal generator associated to a finite-population size process, which corresponds to the form given in Eq.~\eqref{eq:sde_generator}. We consider a continuous-time birth-death process with transition rates $T^{k+}$ and $T^{k-}$ for $0\leq k$. 
Due to the exponentially distributed waiting times, the probability for a single update until time $t$ is $\lambda t\exp(-\lambda t)$, where $\lambda$ is the rate of the corresponding exponential clock. Setting $x=X/N$, the frequency of type-$A$ individuals, we find the infinitesimal generator for the model with finite population size $N$, $\mathscr{G}^N$, to be of the form
\begin{equation}\label{eq:cont_Moran_deriv0}
\begin{aligned}
	(\mathscr{G}^N f)(x) &= \lim_{\Delta t \to 0} \left\{\frac{1}{\Delta t} \left[\underbrace{N T^{xN+} \Delta t e^{-N T^{xN+} \Delta t} \left(f\left(x+\frac{1}{N}\right) - f(x)\right)}_{\text{probability of birth of type } A \text{ until time } \Delta t} \right. \right. \\
	&\qquad \left. \left. + \underbrace{N T^{xN-} \Delta t e^{-N T^{xN-} \Delta t} \left(f\left(x-\frac{1}{N}\right) - f(x)\right)}_{\text{probability of death of type } A \text{ until time } \Delta t} \right] + \underbrace{O(\Delta t)}_{\text{more than one update until time } \Delta t} \right\} \\
	&= N \left( T^{xN +} \left(f\left(x+\frac{1}{N}\right) - f(x)\right) + T^{xN -} \left(f\left(x-\frac{1}{N}\right) - f(x)\right)\right).
\end{aligned}
\end{equation}
We have used the Landau notation $O(\Delta t)$ to summarize processes that scale with order $\Delta t$ or higher.
Doing a Taylor expansion for large $N$ and neglecting the terms of order higher than $1/N^2$ we find
\begin{equation}\label{eq:cont_Moran_deriv}
\begin{aligned}
	(\mathscr{G}^N f)(x) &= N \left( T^{xN+} \left(f\left(x+\frac{1}{N}\right) - f(x)\right) + T^{xN -} \left(f\left(x-\frac{1}{N}\right) - f(x)\right)\right)\\ 
	&\approx N \left( T^{xN +} \left( f(x) + \frac{1}{N}f'(x) + \frac{1}{2 N^2} f''(x) -f(x) \right) \right.\\
	&\qquad \qquad \qquad   \left. + T^{xN -} \left( f(x) - \frac{1}{N}f'(x) + \frac{1}{2 N^2} f''(x) - f(x) \right)  + O\left(\frac{1}{N^3}\right)\right)\\
	&= N \left(\left(T^{xN +} - T^{xN -}\right) \frac{f'(x)}{N}  + \left(T^{xN +} + T^{xN -}\right) \frac{f''(x)}{2N^2}  + O\left(\frac{1}{N^3}\right)\right)\\
	&= \left(T^{xN +} - T^{xN -}\right) f'(x) +\frac{1}{2 N}	\left(T^{xN +} + T^{xN -}\right) f''(x) + O\left(\frac{1}{N^2}\right)\, .
\end{aligned}
\end{equation}
Translating this equation to a stochastic differential equation we identify the single components as 
\begin{equation} \label{eq:moran_sde} 
	\begin{aligned}
		\mu(x) &= \lim_{N\to\infty} \left(T^{xN +} - T^{xN -} \right) \quad \text{ and } \quad 
		\sigma^2(x) = \lim_{N\to\infty} \frac{\left(T^{xN +} + T^{xN -} \right)}{N}.
	\end{aligned}
\end{equation}

Note, that we have made no assumption on the dependence of the transition probabilities on the frequency $x$,
such that this approach is applicable for constant selection, linear frequency dependence arising in 
two player games \citep{traulsen:PRL:2005} or multiplayer games with polynomial frequency dependence \citep{gokhale:PNAS:2010,pena:JTB:2014}.
\begin{conclusion}
	For time-continuous finite population size models with jumps of $\pm 1$, i.e. a birth-death process, the terms of the continuum limit can be computed by Eq.~\eqref{eq:moran_sde}. 
\end{conclusion}

\subsubsection{Example: Moran process with selection and mutation}
Returning to our proto-type processes, we explicitly derive the stochastic differential equation corresponding to the Moran model with selection and mutation. We decouple reproduction and mutation processes, but similar derivations can be made if we assume a coupling of mutations to reproduction events. The selection coefficient is denoted by $s$ and the mutation rates from type $A$ to $B$ and type $B$ to $A$ are given by $u_{A\to B}$ and $u_{B\to A}$, respectively. Then, the transition rates are
\begin{equation}\label{eq:up}
	T^{k+} = (1+s)k \frac{(N-k)}{N} + u_{B\to A} (N-k), \quad \text{and} \quad T^{k-} = (N-k) \frac{k}{N} + u_{A\to B} k.
\end{equation}
%
Inserting these into Eq.~\eqref{eq:moran_sde} yields
\begin{equation}\label{eq:trans_moran} 
	\begin{aligned}
		\mu(x) &= \lim_{N\to\infty} \left(s k \frac{(N-k)}{N} + u_{B\to A} (N-k) - u_{A\to B} k\right), \\
		\sigma^2(x) &= \lim_{N\to\infty} \left(\frac{(2+s) k \frac{N-k}{N} + u_{B\to A} (N-k) +u_{A\to B} k}{N}\right)\, .
	\end{aligned} 
\end{equation}
Depending on the choice of selection and mutation rates, these equations result in different limits. Typically one is interested in non-trivial limits for these equations, i.e.\ a limit so that not both components equal zero. Often this can be achieved by rescaling time (see Section~\ref{sec:scale_time} below) and/or defining the strength of selection and mutation in terms of the population size $N$. As an example, we will focus on two specific limits: (i) strong selection and strong mutation and (ii) weak selection and weak mutation. 

\subsubsection*{Strong selection and mutation}
We consider large selection and mutation rates. We assume that $s$ and $u_i$ do not depend on $N$ but are constant. To obtain a limit equation for the frequency of individuals of type $A$, $x=X/N$, we rescale time by $N$, i.e. $t\mapsto Nt$, which transforms Eq.~\eqref{eq:trans_moran} to
\begin{equation}\label{eq:Moran_strong}
	\begin{aligned}
	\mu(x) &= \lim_{N\to\infty} \left(s x (1-x) + u_{B\to A} (1-x) - u_{A\to B} x \right)=s x (1-x) + u_{B\to A} (1-x) - u_{A\to B} x, \\
		\sigma^2(x) &= \lim_{N\to\infty} \left(\frac{2 x(1-x) + s x(1-x) + u_{B\to A} (1-x) + u_{A\to B} x}{N}\right) = 0\, .
\end{aligned} 
\end{equation}
The first equation is independent of $N$ and the vanishing variance in the second equation implies that the limit process is deterministic. We find the ordinary differential equation
\begin{equation}\label{eq:deterministic}
	dx_t = \mu(x_t) dt = s x_t (1-x_t) + u_{B\to A} (1-x_t) - u_{A\to B} x_t\, ,
\end{equation}
which describes the change in allele frequency in a population under strong selection and mutation over time. 

\subsubsection*{Weak selection and mutation}
In contrast to the previous scenario, we now assume that both selection and mutation are weak. We let $s$ and $u_i$ scale inversely with $N$ and define the constants $\alpha = s N$ and $\nu_i = u_i N$. Inserting these into Eq.~\eqref{eq:trans_moran} (and here without rescaling time), yields
\begin{equation}
\begin{aligned}
	\mu(x) &= \lim_{N\to\infty} \left(\alpha x (1-x) + \nu_{B\to A} (1-x) - \nu_{A\to B} x\right)=\alpha x (1-x) + \nu_{B\to A} (1-x) - \nu_{A\to B} x, \\
	\sigma^2(x) &= \lim_{N\to\infty} \left(2 x (1-x) + \frac{\alpha x (1-x) + \nu_{A\to B} (1-x) +\nu_{A\to B} x}{N}\right)= 2 x (1-x)\, ,
\end{aligned}
\end{equation}
which gives the diffusion limit 
\begin{equation}\label{eq:moran-diff}
	dx_t = \left(\alpha x_t (1-x_t) + \nu_{B\to A} (1-x_t) - \nu_{A\to B} x_t\right) dt + \sqrt{2 x_t(1-x_t)}dW_t\, .
\end{equation}%

Note, that compared to the Wright-Fisher diffusion in Eq.~\eqref{eq:wf-diff}, this limit has twice as much variance.

\subsection{Comparing the variance of Moran and Wright-Fisher process}
Comparing the continuum limits derived from a Wright-Fisher model (Eq.~\eqref{eq:wf-diff}) and a Moran model (Eq.~\eqref{eq:moran-diff}), we see that the variance in the latter limit ($2x(1-x)$) is twice as large as in the former ($x(1-x)$). 
This difference is explained by the different sampling schemes in the individual-based description of the model. To see this, we assume no selection, $s=0$, and mutation, $u_{A\to B}=u_{B\to A} = 0$.

In the Wright-Fisher process, individuals are updated by binomial sampling. The variance of this sampling procedure is $N x (1-x)$, where the factor $N$ vanishes by rescaling the time. This gives $\sigma^2(x) =x(1-x)$.


In the Moran model, or more generally for a birth-death process, the variance is computed by the sum of the transition rates, cf. Eq.~\eqref{eq:moran_sde}. In our example both transitions happen at rate $1$, which explains the additional factor $2$. The difference between the variances is therefore a result of the different sampling schemes of the individual based models.

As a consequence of this difference in the variance $\sigma^2(x)$ between the two models, the Moran diffusion limit progresses twice as fast as the Wright-Fisher diffusion limit which can be seen by the scaling property of the Brownian motion. 
In terms of the original discrete processes, this means that $N$ individual jumps, like in the Moran process, accumulate more variance than one update of the whole population, like in the Wright-Fisher process. The sampling therefore determines the variance and consequently the speed of the continuum limit.

\begin{conclusion}
The Moran process, by definition, has the same mean behavior as the Wright-Fisher model. However, its variance in the diffusion limit is twice the variance of the corresponding Wright-Fisher diffusion. This difference arises from the different sampling schemes of the individual-based models.
\end{conclusion}

\subsection{Change of time scales in the derivation of a continuum limit}\label{sec:scale_time}
In the derivation of the continuum limit we have repeatedly rescaled time to obtain a non-trivial limit, e.g. right before Eqs.~\eqref{eq:wf-diff} and~\eqref{eq:Moran_strong}. Rescaling the time speeds up (or slows down) the original process so that the dynamics of interest, e.g. allele frequency changes, become observable. For example, if the dynamics were to be very fast in the original process, we would need to slow down time appropriately to observe the changes of the quantity of interest more gradually. In general, we are free to chose any scaling of time. 
However, it is important to keep in mind the scaling when interpreting results obtained in the limiting process in terms of the original process. Especially so, if one is interested in quantities involving time, e.g. fixation or extinction times. 

\begin{conclusion}
Different assumptions on the model dynamics, e.g. on selection and mutation, can lead to different continuum limits on the population level. To identify parameter combinations that result in a reasonable continuum limit, one needs to study Eq.~\eqref{eq:moran_sde} to match the orders of the scaling parameter. Rescaling time gives an additional degree of freedom when trying to match these orders to obtain a reasonable limit.
\end{conclusion}

\section{Diffusion approximation}\label{sec:finite}
We have seen that if we let the population size $N$ tend to infinity, we can derive a (stochastic) differential equation describing the studied evolutionary or ecological process. A natural question that arises is how these results relate to finite population size models. 
To study this difference between the finite population process and the continuum limit, we consider the logistic growth equation. The transition rates are given by
\begin{equation}
	\begin{aligned}
		T^{j+} = \beta j\quad \text{and} \quad T^{j-} = j \left(\delta + \frac{\gamma (j-1)}{K}\right)\, .
	\end{aligned}	
\end{equation}
Repeating the steps from the previous section with $y=j/K$, we find the following expressions for the infinitesimal mean and variance:
\begin{equation}\label{eq:log_inf}
	\begin{aligned}
		\mu(y) &= \lim_{K\to\infty} \left(\frac{T_{y K}^+ - T_{y K}^-}{K}\right) = (\beta-\delta)y\left(1-\frac{\gamma y}{(\beta-\delta)}\right), \\
		\sigma^2(y) &= \lim_{K\to\infty} \left(\frac{T_{y K}^+ + T_{y K}^-}{K^2}\right) = \lim_{K\to\infty} \frac{(\beta+\delta+\gamma y)y}{K} = 0\, .
	\end{aligned}
\end{equation}
Thus, the classical deterministic logistic equation is obtained in the infinite population size limit, $K\to\infty$:
\begin{equation}\label{eq:log_det}
	dy_t = (\beta-\delta)y_t\left(1-\frac{\gamma y_t}{\beta-\delta}\right)dt = \mu(y_t) dt\, .
\end{equation}

How well does the finite population size description approximate this deterministic limit? 
One way to approach this question is to simply not take the limit of $K$ to infinity. The finite population size logistic equation, derived from Eq.~\eqref{eq:log_inf}, is then approximated by 
\begin{equation}\label{eq:sde_main}
	dy_t^K = \mu(y_t^K)dt + \sqrt{\frac{(\beta+\delta+\gamma y_t^K)y_t^K}{K}}dW_t\, ,
\end{equation}
where the superscript $K$ in $y_t^K$ indicates the order of magnitude of the carrying capacity.

The approximation in Eq.~\eqref{eq:sde_main} is called \textit{Diffusion approximation}. One can prove formally that this approximation, under the assumptions that the function $\mu(x)$ and $\sigma^2(x)$ are (twice) continuously differentiable, performs equally well as a more accurate analysis based on the central limit theorem \citep[][Theorem 11.3.2]{ethier:book:1986}. For a rigorous discussion of diffusion approximations and their relation to the central limit theorem, see \citet[][Chapter 11]{ethier:book:1986}.

\begin{figure}[t!]
	\centering
	\begin{subfigure}{.5\textwidth}
  		\centering
  		\includegraphics[width=\linewidth]{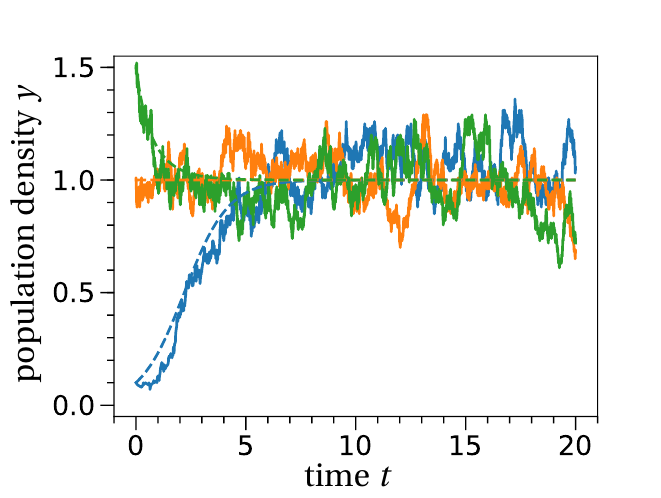}
  		\caption{$K=100$}
	\end{subfigure}%
	\begin{subfigure}{.5\textwidth}
 		\centering
 		\includegraphics[width=\linewidth]{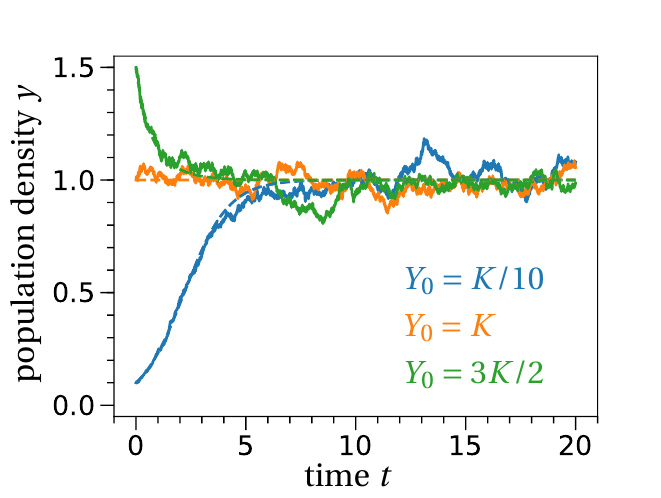}
  		\caption{$K=1,000$}
	\end{subfigure}
	\caption{\textbf{Individual based simulations of the logistic growth model.} 
	(a) For low population sizes, the individual-based simulation (solid lines) fluctuates strongly around the deterministic solution of the population (dashed lines) given by Eq.~\eqref{eq:log_det}. 
	(b) Increasing the scaling parameter $K$, the stochastic fluctuations around the deterministic prediction decrease, until eventually the individual based simulation is indistinguishable from the deterministic curve. The parameter values are chosen as follows: $\beta=2,\delta=1,\gamma=1$. The initial population sizes are stated in subfigure (b).}
	\label{fig:logistic}
\end{figure}

In terms of performance of the diffusion approximation, we see in Figure~\ref{fig:logistic}, that the population size measure $K$ does not need to be very large for the individual-based model to approach the deterministic limit ($K\approx 1,000$ is enough in this example).

\begin{conclusion}
Not taking the limit in Eq.~\eqref{eq:moran_sde} yields the diffusion approximation of the studied model. This approximation is a stochastic differential equation (see Eq.~\eqref{eq:sde_main}) where the variance (typically) scales inversely with the square-root of the scaling parameter.
\end{conclusion}

\section{Stationary distributions}
\begin{figure}[t!]
	\centering
	\begin{subfigure}{.45\textwidth}
  		\centering
  		\includegraphics[height=5.5cm]{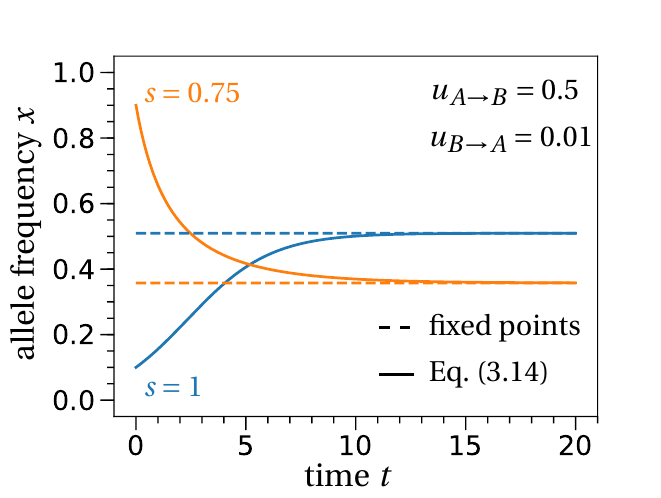}
  		\caption{Deterministic dynamics}
	\end{subfigure}%
	\begin{subfigure}{.55\textwidth}
 		 \centering
 		 \includegraphics[height=5.5cm]{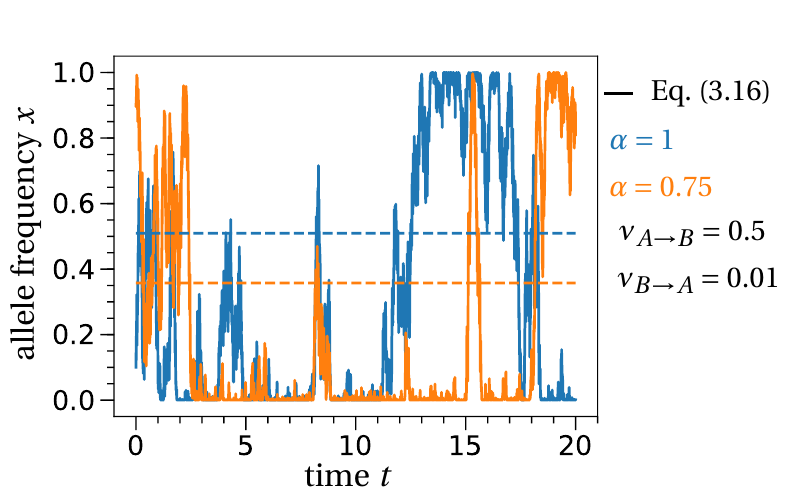}
  		\caption{Stochastic dynamics}
	\end{subfigure}
	\caption{\textbf{Allele frequency dynamics with selection and mutation.} 
	(a) The deterministic system given by Eq.~\eqref{eq:deterministic} converges to the fixed point (dashed line) and remains there. 
	(b) The stochastic process given by Eq.~\eqref{eq:moran-diff} fluctuates strongly in frequency space and spends most time close to the monotypic states $x=0$ and $x=1$.}
	\label{fig:trajectory}
\end{figure}

For the Moran model we have derived two different limits that differ in the assumptions on selection and mutation. If both selection and mutation are strong, the infinite population size limit is a ordinary differential equation. For weak selection and weak mutation we derived a stochastic differential equation. One qualitative difference between these two limits is that trajectories of the deterministic limit will always converge to a fixed point (other limits are possible in general, e.g. limit cycles) while the stochastic differential equation fluctuates indefinitely for positive mutation rates. The deterministic fixed point of the Moran model is given by the solution of Eq.~\eqref{eq:deterministic} equal to zero. 
In our example, a single fixed point $x^\ast$ lies within the interval between $0$ and $1$ and is stable.
Therefore all trajectories will approach this value, e.g. see Figure~\ref{fig:trajectory}(a).

In contrast, Eq.~\eqref{eq:moran-diff} is a stochastic equation. Thus, even if the trajectories approach or hit the deterministic fixed point they will not stay there due to the stochasticity of the Brownian motion, cf. Figure~\ref{fig:trajectory}(b). 
Still, we can make predictions about the time a trajectory spends in certain allele configurations. 
This information is summarized in the stationary distribution, the stochastic equivalent of a deterministic fixed point. If the initial state of the population is given by the stationary distribution, then the distribution of all future time points will not change. 
For birth-death processes, the stationary distribution can be calculated based on detailed balance, i.e. the incoming and outgoing rates need to be equal for every state of the process~\citep{gardiner:book:2004,claussen:PRE:2005,antal:JTB:2009a}.

Formally, the stationary distribution $\psi$ is defined as the solution of
\begin{equation}
	\frac{d}{dt}\mathbb{E}[f(x_t)|x_0\sim \psi] = 0\, ,
\end{equation}  
where $x_0\sim \psi$ denotes that $x_0$ is distributed according to $\psi$ and $f$ is an arbitrary function. Importantly, this condition means that the distribution of allele frequencies does not change over time because its derivative in time is zero (for any choice of $f$). The above equation can be solved with the infinitesimal generator (see \citet[Chapter 3.6]{etheridge:LN:2012}). The solution is expressed in terms of the speed measure density $m(x)$ (see Box~2 for its definition) and is given by
\begin{equation}\label{eq:stationary}
	\psi(x) = \frac{m(x)}{\int_0^1 m(y) dy}\, .
\end{equation}

Intuitively, the speed measure at a point $x$, $m(x)$, quantifies the time which the process spends in this state. Therefore, $\psi(x)$ is nothing but the average time spent in state $x$. 

\begin{figure}[t!]
	\begin{mybox}{Box 2: Scale function and speed measure of a one-dimensional diffusion}
	
		A one-dimensional stochastic diffusion can be transformed into a standard Brownian motion. Since the Brownian motion is well-studied, a lot of results can then be translated to the stochastic diffusion by the transformation functions, the \textit{scale function} and the \textit{speed measure}.
		
		First, we rescale the space of the original process by the scale function. It is defined by 
\begin{boxequation}\label{eq:scale_function}
		S(x) = \int^x \exp \left(-2\int^y \frac{\mu(z)}{\sigma^2(z)}dz\right)dy, \quad x\in (0,1),
\end{boxequation}
where the lower boundaries of the integrals can be chosen arbitrarily. 
The name of this function derives from the fact that for a one-dimensional diffusion $x_t$ satisfying
\begin{boxequation}
	dx_t = \mu(x_t)dt + \sigma(x_t)dW_t, 
\end{boxequation}
the scaled process $S(x_t)$ becomes a time-changed Brownian motion on the interval $[S(0),S(1)]$, i.e. there is no deterministic contribution in the scaled process. The process $S(x_t)$ is a Brownian motion with a `non-standard' time scale. To map this time-changed Brownian motion to the time scale of a standard Brownian motion one needs to rescale time by the speed measure $M$. It defines how much faster (or slower) the process $S(x_t)$ is evolving compared to a standard Brownian motion. The speed measure is defined by
\begin{boxequation}\label{eq:speed_measure}
	M(x) = \int^x m(y) dy,\quad \text{ with }\quad m(y) = \frac{1}{\sigma^2(y)S'(y)}
\end{boxequation}
the density of the speed measure. The time is then rescaled by $\tau(t) = \int_0^t m(S(x_s))ds$.

Compactly written, we have changed the stochastic diffusion $x_t$ to the standard Brownian motion by the following steps:
\[ 
	\begin{array}{ccccc}
		x_t & & S(x_t) = B_t & & B_{\tau(t)}\\
		\text{(stochastic} & \smash{\raisebox{.5\normalbaselineskip}{$\xrightarrow{x\mapsto S(x)}$}} & \text{(time-changed }	& \smash{\raisebox{.5\normalbaselineskip}{$\xrightarrow{t\mapsto \tau(t)}$}} & \text{(standard} \\
		\text{diffusion)}& & \text{Brownian motion)} & & \text{Brownian motion)}\end{array}\, .
\]

	\end{mybox}
\end{figure}
\begin{conclusion}
	The stationary distribution of a one-dimensional diffusion can be expressed in terms of the density of the speed measure $m(x)$ through Eq.~\eqref{eq:stationary}. The density of the speed measure is given by the scale function corresponding to the stochastic diffusion process, Eqs.~\eqref{eq:scale_function} and \eqref{eq:speed_measure} in Box 2.
\end{conclusion}

\subsection{Stationary distribution of the Wright-Fisher diffusion}
As an example let us consider the Wright-Fisher diffusion with selection and mutation as derived in Eq.~\eqref{eq:wf-diff} (and Eq.~\eqref{eq:moran-diff} when derived from the Moran model), i.e.
\begin{equation}\label{eq:true_wf}
	dx_t = \left(\alpha x_t (1-x_t) + \nu_{B\to A} (1-x_t) - \nu_{A\to B} x_t\right) dt + \sqrt{x_t(1-x_t)}dW_t\, . 
\end{equation} 
Computing Eq.~\eqref{eq:stationary} with help of the quantities defined in Box~2, one obtains
\begin{equation}\label{eq:stat_wf}
	\psi(x) =  e^{2\alpha x}x^{2\nu_{B\to A} -1}(1-x)^{2\nu_{A\to B} -1} \frac{\Gamma (2(\nu_{A\to B}+\nu_{B\to A}))}{\Gamma (2\nu_{A\to B}) \Gamma (2\nu_{B\to A}){}_1F_1 (2\nu_{B\to A},2(\nu_{A\to B}+\nu_{B\to A}),\alpha)}\, ,
\end{equation}
where $\Gamma (x)$ is the Gamma function and ${}_1 F_1(a,b,z)$ is the generalized hypergeometric function.

\begin{figure}[t]
	\center
	\includegraphics[width=0.75\textwidth]{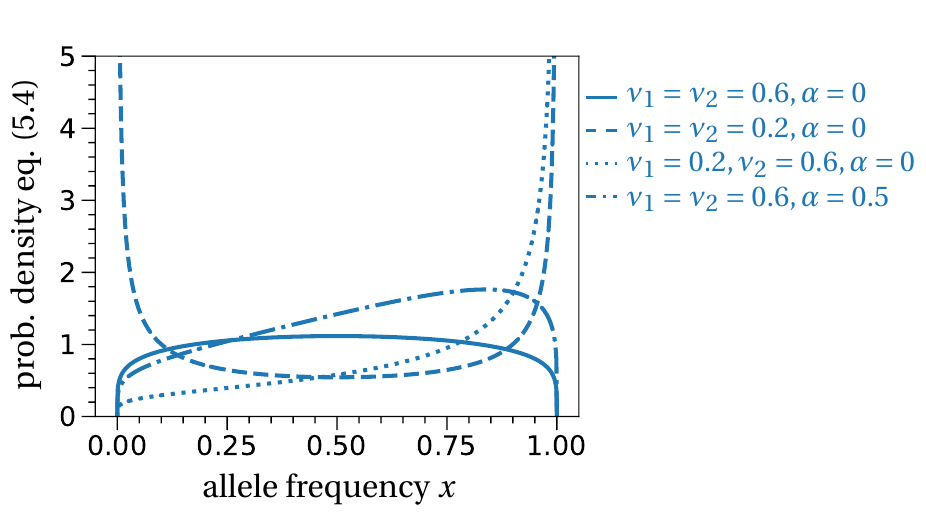}
	\caption{\textbf{Stationary distribution of the Wright-Fisher diffusion with selection and mutation.} The lines are given by Eq.~\eqref{eq:stat_wf}. Larger mutation rates accumulate more probability on intermediate allele frequencies (compare solid and dashed lines). Selection (or asymmetric mutation) skews the stationary distribution towards the selectively favored type (or type with the lower mutation rate), see the dash-dotted (dotted) line.}
	\label{fig:stat}
\end{figure}

The equation itself does not provide much insight. To illustrate the possible shapes of stationary distributions, we plot several choices of mutation rates and selection coefficients in Fig.~\ref{fig:stat}. We see that for higher mutation rates, more probability mass is allocated to intermediate allele frequencies (compare the solid and dashed lines). In this case, the Wright-Fisher diffusion spends more time in states of coexistence than in monotypic states (the boundaries of the allele frequency space in Fig.~\ref{fig:stat}) because temporary extinction events are prevented by recurrent mutations. If the mutation rates are asymmetric (dotted line), the stationary distribution is skewed towards the type with the lower mutation rate. If one type is favored selectively, dash-dotted line, the stationary distribution is skewed towards the favored type.

Stationary distributions are the stochastic equivalents of deterministic fixed points and as such provide a basic description and a starting point for further analysis of the qualitative behavior of a stochastic model, especially in situations where polymorphisms of alleles, coexistence of species, or spatial population distributions are to be expected \citep[e.g.][]{polansky:TPB:1979,turelli:JMB:1981,gaston:PRSB:2002,lehmann:JEB:2012,czuppon:JEB:2019}. 

\subsection{Quasi-stationary distribution of the logistic process}
Next, we consider the diffusion approximation of the logistic growth model, i.e.
\begin{equation}\label{eq:log_finite}
	dy_t^K = (\beta-\delta-\gamma y_t^K)y_t^K dt + \sqrt{\frac{(\beta+\delta+\gamma y_t^K)y_t^K}{K}}dW_t\, .
\end{equation} 
The logistic process for finite $K$ has a (unique) absorbing state, $y=0$ because there is no transition from this state to positive population densities. Once the population went extinct, it remains so. Since the extinction state is accessible from all values $y>0$, the population will go extinct with probability $1$. The only stationary distribution is the point-measure on $0$, i.e. $\psi(y) = \delta_0$ ($\delta_x$ is the Dirac measure at point $x$).

In contrast, the positive deterministic population equilibrium, $y^\ast = (\beta-\delta)/\gamma$, is a stable fixed point of the deterministic system. Considering large values of the deterministic equilibrium ($K\gg 1$), we expect the finite population size process from Eq.~\eqref{eq:log_finite} to remain close to this value for long times. In fact, the expected extinction time of the logistic growth process when started in the positive population equilibrium is of order $\exp(K)$ \citep{champagnat:SP:2006}. This suggests that the process will be in a quasi-stationary state, i.e., before its extinction the population is described by the stationary distribution of the corresponding logistic process conditioned on non-extinction.

Formally, the quasi-stationary distribution is computed by conditioning the original process on its survival. This means that the transition rates change and the novel process can be analyzed by the techniques described above. However, this method goes beyond the scope of this manuscript. For a theoretical treatment of this topic in the context of the logistic equation we refer to \citet{cattiaux:AnAP:2009,assaf:JSM:2010,meleard:ProbS:2012}. 
For a general review on methods related to quasi-stationary distributions see \citet{ovaskainen:TREE:2010}. 

Another way to approximate the quasi-stationary distribution when extinction is very unlikely for long times (which is the case for large $K$), is provided by the central limit theorem (sometimes also called linear noise approximation). Here, the distribution of the process is derived from its local dynamics around the deterministic fixed point $y^\ast$ \citep{ethier:book:1986,van_kampen:book:2007}. The underlying assumption is that the population stays close to its positive steady state and just slightly fluctuates around this value. This is only a valid assumption when the probability of extinction within the studied time-frame is essentially zero. These small fluctuations are described by a Gaussian distribution. Formally this translates to 
\begin{equation}\label{eq:lna}
	y_t^K \approx y_t + \frac{1}{\sqrt{K}} U,
\end{equation} 
where $U$ is a Gaussian random variable and $y_t$ the deterministic trajectory. Writing $\mu(y) = (\beta-\delta-\gamma y)y$ and $\sigma^2(y) = (\beta+\delta+\gamma y)y$, the dynamics of $U$ can be rewritten as
\begin{equation}\label{eq:ou_approx}
	\begin{aligned}
		\frac{1}{\sqrt{K}}dU &\approx dy_t^K -dy_t = \left(\mu(y_t^K)-\mu(y_t)\right)dt +\sqrt{\frac{\sigma^2(y_t^K)}{K}}dW_t \\
		&\approx \mu'(y_t) (y_t^K-y_t)dt + \sqrt{\frac{\sigma^2(y_t^K)}{K}}dW_t \qquad \text{(Taylor series approximation)}\\
		&\approx \frac{1}{\sqrt{K}}\left(\mu'(y_t) U dt + \sigma(y_t^K) dW_t\right)\, .
	\end{aligned}
\end{equation}
Evaluating the process $U$ at $y_t^K =y_t= y^\ast$ we obtain a description of the variance in the deterministic fixed point. For this fixed choice of $y_t$ and $y_t^K$, $U$ becomes a Ornstein-Uhlenbeck process. Its stationary distribution is given by (cf. Eq.~\eqref{eq:ou_stat} in Appendix~\ref{app:ou_process})
\begin{equation}
	\psi^U(y) \sim \mathcal{N}\left(0,-\frac{\sigma^2(y)}{2 \mu'(y)}\right)\, . 
\end{equation}

This distribution describes the fluctuations of the process $y_t^K$ around the deterministic steady state~$y^\ast$. Therefore, when plugging the distribution $\psi^U$ into the original process from Eq.~\eqref{eq:lna}, we find the quasi-stationary distribution of $y_t^K$ around the deterministic equilibrium $y^\ast$ which is given by
\begin{equation}\label{eq:stat_log}
	\psi(y) \sim \mathcal{N}\left(y^\ast,-\frac{\sigma^2(y)}{2 K \mu'(y)}\right)\, .
\end{equation} 
We see that for increasing population sizes $K$, the variance is decreasing and vanishes in the limit $K\to\infty$. 
%

\begin{conclusion}
	If the deterministic process has a stable steady state but is almost surely going extinct for finite population sizes, a quasi-stationary distribution can be computed to describe the behavior of the process conditioned on survival. If the extinction probability is very low, an approximation of this distribution is given by the linear noise approximation where the variance around the deterministic steady state is modeled by the Ornstein-Uhlenbeck process derived from Eq.~\eqref{eq:ou_approx}.
\end{conclusion}	

\section{Fixation probabilities}
We have seen that stochastic descriptions of processes can lead to outcomes that are different from their deterministic counterparts. Here, we study one of these phenomena, the probability for a certain type to become fixed in a population. For one-dimensional stochastic differential equations fixation probabilities can be computed explicitly. As before, we denote by $x_t$ the frequency of type $A$ individuals at time $t\geq 0$ in the population. Using the fact that we can transform a one-dimensional diffusion into a time-changed Brownian motion through the scale function $S(x)$ in Eq.~\eqref{eq:scale_function}, the computation of the fixation probability simplifies to an argument about hitting probabilities of a Brownian motion. With $p_{\text{fix}}(x_0)=\mathbb{P}(x_{\infty} = 1|x_0)=\mathbb{P}_{x_0}(x_{\infty} = 1)$ we find
\begin{equation}\label{eq:fix_prob}
	S(x_0) = \mathbb{E}[S(x_t)|x_0] \stackrel{t \gg 1}{=} p_{\text{fix}}(x_0) S(1) + (1-p_{\text{fix}}(x_0)) S(0) \quad \Leftrightarrow \quad p_{\text{fix}}(x_0) = \frac{S(x_0)-S(0)}{S(1)-S(0)}, 
\end{equation}
where we have used that the mean of the Brownian motion does not change over time and remains at its conditioned, here also initial, value $x_0$ (first equality). Stochastic processes with this property are called \textit{martingales}. The second equality is explained by the process being absorbed at one of the two boundaries $x=0$ or $x=1$ at large times. For a formal derivation we refer to \citet[Chapter 15.3.3]{otto:book:2007} or \citet[Lemma 3.14]{etheridge:LN:2012}. 

As an example, let us consider the Wright-Fisher diffusion with selection and without mutations ($\nu_{A\to B}=\nu_{B\to A}=0$) given in Eq.~\eqref{eq:true_wf}. We have $\mu(x)=\alpha x(1-x)$ and $\sigma^2(x) = x(1-x)$ such that the scale function simplifies to
\begin{equation} 
	S(x) = \int^x \exp\left(-2 \int^y \frac{\alpha z(1-z)}{z(1-z)}dz\right)dy = -\frac{1}{2\alpha} \exp(-2\alpha x). 
\end{equation}
Recalling the definition of $\alpha = s N$ for a finite population of size $N$ and plugging this into Eq.~\eqref{eq:fix_prob} yields
\begin{equation}\label{eq:fix_wf}
	\begin{aligned}
	\mathbb{P}_{x_0}(x_{\infty} = 1) &= \; \; \frac{1-e^{-2\alpha x_0}}{1-e^{-2\alpha}} = \frac{1-e^{-2s x_0 N}}{1-e^{-2sN}}\\
	&\stackrel{\mathclap{Ns\gg 1}}{\approx} \; \; \; \; 1-e^{-2s x_0 N} \\
	&\stackrel{\mathclap{s\ll 1}}{\approx} \; \; \; \; 2sx_0 N, 
\end{aligned} 
\end{equation}
which for $x_0=1/N$ becomes $\mathbb{P}_{1/N}(x_{\infty} = 1)\approx 2s$, the result of Haldane for the fixation of a single mutant copy in a population of size $N$ \citep{haldane:PCPS:1927}. The first line of Eq.~\eqref{eq:fix_wf}, the classical result of fixation probabilities when derived from diffusion theory, and its applicability has been subject to extensive research see e.g. \citet{buerger:JMB:1995} and references therein; for a more general review on fixation probabilities we refer to \citet{patwa:JRSI:2008}.

Of course, the fixation probability can also be calculated for more complicated stochastic differential equations where the sign of the deterministic dynamics $\mu(x)$ depends on the population configuration. Most classically, these frequency-dependent problems were studied in deterministic evolutionary game theory introduced by \citet{maynard-smith:Nature:1973} (see also \citet{hofbauer:book:1998} for an introduction to evolutionary game dynamics). In Appendix~\ref{app:freq} we (re-)derive the fixation probability in case of frequency-dependent selection.

\begin{conclusion}
The fixation probability of a one-dimensional diffusion is given by the scale function as stated in Eq.~\eqref{eq:fix_prob}.
\end{conclusion}

\section{Mean time to fixation}
A related quantity of interest is the expected time to fixation (or extinction from the other type's point of view), i.e. the average time of coexistence of two types. Again, the calculation relies on a special function, this time \textit{Green's function} $G(x,y)$, 
which can be interpreted as the average time that a diffusion started in $x$ spends in the interval $[y,y+dy)$ before reaching one of the boundaries~\citep[Chapter 3.5][]{etheridge:LN:2012}. It is therefore also called sojourn time density \citep{ewens:book:2004}. 
It is defined as

\begin{equation}\label{eq:Green}
	G(x,y) = \left\{ \begin{array}{ll} 2\frac{S(x)-S(0)}{S(1)-S(0)} (S(1)-S(y))m(y), & 0\leq x\leq y\leq 1, \\ 2 \frac{S(1)-S(x)}{S(1)-S(0)}(S(y)-S(0))m(y), & 0\leq y\leq x\leq 1, \end{array} \right. 
\end{equation}
where $S(x)$ is the previously defined scale function and $m(x)$ denotes the speed measure density (see Box 2).  

The expected time to fixation for a process started at frequency $x$, denoted $\mathbb{E}_x[\tau]$, is then given by (see \citet[Section 4.4]{ewens:book:2004}) 
\begin{equation}\label{eq:ext_time}
	\begin{aligned}
		\mathbb{E}_x[\tau] &= \int_0^1 G(x,y) dy \\
	\end{aligned}
\end{equation}
This corresponds to the summation of the sojourn times in the discrete case, see e.g. \citet{ohtsuki:JTB:2007c} for an application in finite populations.
In some cases, the result of this equation yields a analytically tractable result, e.g. for the neutral Wright-Fisher diffusion
\begin{equation}
	dx_t = \sqrt{x_t(1-x_t)}dW_t\,.
\end{equation} 
In this case, the scale function and speed measure density are given by
\begin{equation} 
	S(x) = x \quad \text{and}\quad m(x) = \frac{1}{x(1-x)}. 
\end{equation}
Then, the expected time to fixation of one of the two alleles can then be expressed as
\begin{equation}
	\begin{aligned}
		\mathbb{E}_x[\tau] &= \int_0^x 2 (1-x)\frac{y}{y(1-y)}dy + \int_x^1 2x\frac{(1-y)}{y(1-y)}dy\\
		&= 2(1-x)\ln((1-x)^{-1}) + 2x\ln(x^{-1}).
	\end{aligned} 
\end{equation}
In Appendix~\ref{app:freq} we consider the more involved example of frequency-dependent selection \citep{altrock:NJP:2009, pfaffelhuber:TPB:2018}. 

Similar to fixation probabilities, the mean time to fixation has been studied extensively through stochastic diffusions, see \cite{kimura:Genetics:1969} for an early reference. It is especially important in population genetics where one is interested in the time to extinction or fixation of newly arising alleles, e.g. \citep{vanHerwaarden:TPB:2002}. On the macroscopic scale, the mean time to extinction or fixation is for example applied in the context of population extinction \citep{lande:evolution:1994} and speciation events \citep{yamaguchi:JRSIF:2013}.

\begin{conclusion}
Expected unconditional fixation times, i.e. the expected time of coexistence of two alleles in a population, can be calculated by integrating over Green's function (the mean occupation time of a certain frequency until extinction), as shown in Eq.~\eqref{eq:ext_time}.
\end{conclusion}

\section{Discussion and Conclusion}

We have outlined how to derive a stochastic differential equation from an individual-based description of two classical models in evolutionary theory and theoretical ecology, the Wright-Fisher diffusion and the logistic growth equation. 
The resulting stochastic differential equation in one dimension describes the evolution of the allele frequency or population density under study, respectively. Using probabilistic properties of this equation, i.e. transforming it to a standard Brownian motion (Box 2), it is possible to analytically derive the (quasi-) stationary distribution, fixation probability and the mean time to fixation. As an example, we derived these quantities for the Wright-Fisher diffusion.

The diffusion process emerges as the infinite population size limit. However, as we have seen in Section~\ref{sec:finite}, one can also derive a finite population size approximation of the dynamics, the diffusion approximation. The fixation probability, mean extinction time and stationary distribution are accessible by the same means as for the continuum limit. Applications of diffusion approximations are abundant and cover diverse topics \citep[e.g.][]{traulsen:PRL:2005,reichenbach:PRL:2007,assaf:JTB:2011,houchmandzadeh:BioSys:2015,constable:PNAS:2016,debarre:TPB:2016,kang:JPA:2017,koopmann:TPB:2017,serrao:JPA:2017,czuppon:TPB:2018,czuppon:JMB:2018,parsons:JRSI:2018,mcleod:Evolution:2019,schenk:BMCEvolBio:2020}. 

Apart from the fixation probability and the mean time to fixation, the (quasi-)stationary distribution is a commonly used measure to describe stochastic processes. Its calculation through the speed measure of the associated scaled process (Box 2) is (in many cases) numerically straightforward. If the process has an absorbing state, e.g. an extinction boundary of the population, the stationary distribution is not meaningful. Here, the quasi-stationary distribution describes the stationary distribution conditioned on the survival of the population. For negligible extinction probabilities, i.e. very large survival probabilities of the population, the functional central limit theorem (or linear noise approximation) can be used to approximate this quasi-stationary distribution.
In the theoretical biology literature, this method is frequently used in models of gene regulatory networks (see \citet[][]{anderson:book:2015} for a mathematical introduction), and less so in the context of ecology or evolution (e.g. \citet{boettiger:TPB:2010,kopp:JMB:2018,wienand:JRSI:2018,czuppon:Genetics:2019}; and \citet{assaf:JPA:2017} for a review of the physics literature related to this topic). 

Lastly, we did not cover multi-dimensional or spatially explicit stochastic differential equations in this methods review. These processes are often much more complicated to analyze. Here, we aimed to give a basic introduction into the derivation of a continuum limit from an individual-based model. We hope, that with our basic comparisons between different approaches used in different subfields of theoretical and mathematical biology, we help newcomers in the field to get more familiar with these methods. 

\subsection*{Acknowledgments}
We are very grateful to Florence D\'{e}barre who carefully read an earlier version of the manu\-script and made numerous suggestions that led to the current version. Both authors appreciate generous funding from the Max Planck Society. PC also received funding from the Agence Nationale de la Recherche, grant number ANR-14-ACHN-0003 provided to Florence D\'{e}barre and the European Union's Horizon 2020 research and innovation program under the Marie Sk{\l}odowska-Curie grant agreement PolyPath 844369.

\subsection*{Author contribution}
Both authors conceived and designed the methods review, wrote and critically revised the manuscript, and gave final approval to publication.

\bibliographystyle{plainnat}
\bibliography{hiker.bib}

\newpage
\begin{appendix}
\renewcommand\thesection{S\arabic{section}}
%
\section{Deriving a stochastic differential equation from the Wright-Fisher model with selection and mutation}\label{app:wf_selection}
In the main text we have derived the Wright-Fisher diffusion in the absence of selection and mutation. Here, we provide the calculation steps when including both these processes.

We say that type $A$ alleles are beneficial (deleterious) if $s>0$ ($s<0$). Given that there are $k$ type $A$ individuals in the population, the probability for an offspring to choose a type $A$ individual as a parent is given by 
\begin{equation} 
	p_k = \frac{(1+s)k}{(1+s)k + N-k}\ .
\end{equation} 
We can also add mutations to the Wright-Fisher model, i.e. type $A$ individuals can mutate to type $B$ and vice versa. We set $u_{A\to B}$ as the probability to mutate from type $A$ to $B$ and $u_{B\to A}$ as the mutation probability from $B$ to $A$. Then the probability for an individual to be of type $A$ given $k$ type $A$ individuals in the parental generation reads
\begin{equation} 
	p_k = \frac{(1+s)k(1-u_{A\to B})}{(1+s)k+N-k}+ \frac{u_{B\to A}(N-k)}{(1+s)k + N-k}.
\end{equation}
In this model mutation is intimately connected with the reproduction mechanism. For the Moran model, compare Section~\ref{sec:infinite_limit}, these processes do not necessarily need to be coupled (even though this would, biologically speaking, make the most sense).

Following the same methodology as for the neutral Wright-Fisher model in the main text, we can derive a diffusion process by computing the infinitesimal mean and variance. Writing $x_t = X_t/N$ and setting $\Delta t = 1/N$, we obtain for the infinitesimal change in allele frequency
\begin{equation} 
	\begin{aligned}
		\frac{1}{\Delta t} \mathbb{E}[x_{\Delta t}-x_0 | x_0=x] &= \mathbb{E}[X_{\Delta t}-X_0 | X_0=k] = (N p_k - k)\\
	&= \left(N\frac{(1+s)k(1-u_{A\to B})}{N+sk}+ N\frac{u_{B\to A}(N-k)}{N+sk} - k\right)\ .
	\end{aligned} 
\end{equation}
This is a rather unhandy expression. However, we can make further progress by assuming that selection and mutation are weak, i.e. we set $s=\alpha/N$ and $u_i=\nu_i/N$. Rewriting the equation in terms of $\alpha$ and $\nu_i$, expanding the equation in terms of $1/N$, and neglecting terms of order $1/N^3$ and higher we find
\begin{equation} 
	\begin{aligned}
	\frac{1}{\Delta t}\mathbb{E}[x_{\Delta t}-x_0 | x_0=x] &= N \frac{(1+\frac{\alpha}{N})k(1-\frac{\nu_{A\to B}}{N})}{N+\frac{k \alpha}{N}}+ N \frac{\frac{\nu_{B\to A}}{N}(N-k)}{N+\frac{k\alpha}{N}} - k\\
	&= k + \frac{\alpha k}{N} - \frac{\nu_{A\to B} k}{N} + \frac{\nu_{B\to A} (N -  k)}{N} - k - \alpha\frac{k^2}{N^2}+ O\left(\frac{1}{N^3}\right)\\
	&= \alpha x\left(1-x\right) - \nu_{A\to B} x + \nu_{B\to A} \left(1-x\right) + O\left(\frac{1}{N^3}\right).
	\end{aligned}
\end{equation}
Thus, for the infinitesimal mean in the infinite population size limit we find
\begin{equation}
	\mu(x) = \alpha x(1-x) - \nu_{A\to B} x + \nu_{B\to A} (1-x). 
\end{equation}
The infinitesimal variance in terms of $\alpha$ and $\nu_i$ derives to
\begin{equation}\label{eq:wf_var}
	\begin{aligned} 
	\frac{1}{\Delta t} \mathbb{E}\left[(\Delta x_t)^2|x_0=x\right] &= \frac{1}{N}\bigg(\mathbb{E}[X_{t+\frac{1}{N}}^2 - 2X_t X_{t+\frac{1}{N}} + X_t^2 | X_0=k]  \bigg)\\
	&= \frac{1}{N}\bigg(\mathbb{V}[X_{t+\frac{1}{N}}| X_0=k] + \left(\mathbb{E}[X_{t+\frac{1}{N}}| X_0=k] - k \right)^2\bigg) \\
	&= p_k(1-p_k) + \frac{1}{N} \mu^2(x)\\
	&=  x (1-x) + O\left(\frac{1}{N}\right),
	\end{aligned}
\end{equation}
where we used 
\begin{equation} 
	p_k = \frac{\left( k +\frac{k \alpha}{N} - \frac{k\nu_{A\to B}}{N} - \frac{k\alpha\nu_{A\to B}}{N^2}+\frac{\nu_{B\to A}}{N}(N-k)\right)}{N+\frac{\alpha k}{N}} = x + O\left(\frac{1}{N}\right). 
\end{equation}
Therefore, the infinitesimal variance for $N\to \infty$ is given by
\begin{equation}
	\sigma^2(x) = x(1-x).
\end{equation}
Putting together the final results for the infinitesimal mean and variance we get the weak selection and mutation limit of the Wright-Fisher model with selection and mutation, i.e.
\begin{equation}\label{eq:wf_sel}
	dx_t = \big(  \alpha x_t(1-x_t) - \nu_{A\to B} x_t + \nu_{B\to A} (1-x_t)\big) dt + \sqrt{x_t(1-x_t)}dW_t\ .
\end{equation}
\clearpage

\section{Infinitesimal mean and variance}\label{app:sde}
Given the stochastic differential equation
\begin{equation}\label{eq:box_sde}
	dx_t = \mu(x_t)dt + \sigma(x_t)dW_t, 
\end{equation}
the corresponding infinitesimal generator is defined by
\begin{equation}\label{eq:gen_app}
	(\mathscr{G}f)(x) = \mu(x) f'(x) + \frac{1}{2}\sigma^2(x) f''(x).
\end{equation}
The connection between the infinitesimal generator and its associated stochastic differential equation is outlined in more detail in e.g. \citet[][Chapter 23]{kallenberg:book:2002}. Briefly, one needs to apply It\^{o}'s formula \citep[][Theorem 17.18]{kallenberg:book:2002} to the process $f(x_t)$, where $x_t$ solves the stochastic differential equation in Eq.~\eqref{eq:box_sde}: 
\begin{equation}
	\begin{aligned}
	df(x_t)\; &\stackrel{\text{It\^o}}{=}\; f'(x_t) dx_t + \frac{\sigma^2(x_t)}{2}f''(x_t) dt \quad \stackrel{\mathclap{\text{Eq.}~\eqref{eq:box_sde}}}{=}\quad \left(\mu(x_t) f'(x_t) + \frac{\sigma^2(x_t)}{2} f''(x_t)\right) dt + \sigma(x_t) dW_t\, .
	\end{aligned}
\end{equation}
Taking the expectation yields the result since the last term on the right hand side vanishes (the mean of a standard Brownian motion is zero).\\

Additionally, we show that $\mu(x)$ is indeed the infinitesimal mean of the stochastic process with infinitesimal generator $\mathscr{G}$ as given in Eq.~\eqref{eq:gen_app}. Setting $f_1(x)=x$ (and thus $f_1''(x) = 0$) we find for the infinitesimal change of the mean
\begin{equation}\label{eq:inf_exp}
	\lim_{\Delta t\to 0} \frac{1}{\Delta t} \mathbb{E}[x_{\Delta t}-x_0|x_0=x] \stackrel{\text{def.}}{=} (\mathscr{G}f_1)(x) \stackrel{\text{Eq.}~\eqref{eq:gen_app}}{=} \mu(x).
\end{equation}
Similarly, we see that $\sigma^2(x)$ is the infinitesimal variance. With $f_2(x) = x^2$ we have
\begin{equation}\label{eq:inf_var}
	\begin{aligned}
		\lim_{\Delta t\to 0} \frac{1}{\Delta t} \mathbb{V}[(x_{\Delta t}-x_0)|x_0=x]\; \; &\stackrel{\mathclap{\text{def.}}}{=} \; \; \lim_{\Delta t\to 0} \frac{1}{\Delta t} \mathbb{E}[(x_{\Delta t}-x_0-\mathbb{E}[x_{\Delta t}-x_0|x_0=x])^2|x_0=x]\\
		&=  \; \; \lim_{\Delta t\to 0} \frac{1}{\Delta t} \left(\mathbb{E}[(x_{\Delta t}-x_0)^2|x_0=x] - \underbrace{\left(\mathbb{E}[x_{\Delta t}-x_0|x_0=x]\right)^2}_{=O((\Delta t)^2)}\right)\\
		&= \; \; \lim_{\Delta t\to 0} \frac{1}{\Delta t} \mathbb{E}[x_{\Delta t}^2 - x_0^2 -2 x_0(x_{\Delta t}-x_0) | x_0 = x]\\
			&\stackrel{\mathclap{\text{def.}}}{=}  \; \; (\mathscr{G}f_2)(x) - 2x (\mathscr{G}f_1)(x) \\		
			&\stackrel{\mathclap{\text{Eq.}~\eqref{eq:gen_app}}}{=} \; \; \; \; \; 2x \mu(x) + \sigma^2(x) - 2 x \mu(x) \; \; = \; \; \sigma^2 (x).
	\end{aligned} 
\end{equation}
This justifies that calculating the infinitesimal mean and variance (right hand sides in Eq.~\eqref{eq:inf_mean_var}), indeed yields the functions $\mu$ and $\sigma^2$ of the diffusion.

\clearpage

\section{Ornstein-Uhlenbeck process}
\label{app:ou_process}
The differential equation that defines an Ornstein-Uhlenbeck process \citep{uhlenbeck:PhysicalReview:1930}, the only stationary Gaussian Markov process, is given by
\begin{equation}
	dx_t = \omega(\mu-x_t) dt + \sigma dW_t,
\end{equation}
where $\mu$ is the expectation in stationarity, $\sigma$ the standard deviation and $\omega$ 
the speed of approaching the value $\mu$. This stochastic differential equation defines a Gaussian process, i.e. a stochastic process with independent and normally distributed increments. It can be solved exactly, where the dynamics of the mean is given by
\begin{equation}
	\mathbb{E}[x_t] = x_0 e^{-\omega t} + \mu\left(1-e^{-\omega t}\right),
\end{equation}
and the dynamics of the covariance by
\begin{equation}
	\mathbf{cov}(x_s,x_t) = \mathbb{E}[(x_s-\mathbb{E}[x_s])(x_t-\mathbb{E}[x_t])]=\frac{\sigma^2}{2\omega} \left(e^{-\omega|t-s|} - e^{-\omega(t+s)}\right)\, . 
\end{equation}
From this we can read off the stationary distribution of the Ornstein-Uhlenbeck process by letting $t$ tend to infinity. It is distributed normally as 
\begin{equation}\label{eq:ou_stat}
	\psi(x) \sim \mathcal{N}\left(\mu,\frac{\sigma^2}{2\omega}\right)\, .
\end{equation}

\clearpage
\section{Frequency-dependent selection}\label{app:freq}
In the main text we have exclusively considered situations with constant selection coefficients $s$ (or $\alpha$). Here instead, we apply the derived formulas for the fixation probability and the mean time to fixation for the case of frequency-dependent selection. More precisely, we consider a stochastic diffusion with linear frequency dependence (see \citet{traulsen:PRE:2006c} for a physical formulation of this and \citet{pfaffelhuber:TPB:2018} for a more general mathematical analysis). 
We denote the strength of selection by $\alpha$ and let $u,v$ be arbitrary real numbers. We write $\alpha x (1-x)(u x + v)$ for the linear frequency-dependent dynamics of selection.
Then, the allele frequency evolves according to the following equation
\begin{equation} 
	dx_t = \alpha x_t(1-x_t)(u x_t + v) dt + \sqrt{x_t(1-x_t)}dW_t.
\end{equation}
We have $\mu(x) = \alpha x(1-x)(u x+ v)$ and $\sigma^2(x) = x(1-x)$. 

\subsubsection*{Fixation probability}
Recall the formula for the fixation probability, Eq.~\eqref{eq:fix_prob}
\begin{equation}\label{eq:rep_fix}
	\mathbb{P}_{x_0}(x_{\infty} = 1) = \frac{S(x_0)-S(0)}{S(1)-S(0)},
\end{equation}
where $S(x)$ is the scale-function given in Eq.~\eqref{eq:scale_function} and given by
\begin{equation}\label{eq:rep_scale}
	S(x) = \int^x \exp \left(-2\int^y \frac{\mu(z)}{\sigma^2(z)}dz\right)dy, \quad x\in (0,1).
\end{equation}

For $\alpha \ll 1$, we can linearize the exponential and write the scale function as
\begin{equation}\label{eq:freq_scale}
	\begin{aligned} 
		S(x) &= \int^x \exp\left(-2 \int^y \alpha (v + u z)dz\right)dy \\
		&= \int^x \exp\left( -2 \alpha vy - \alpha uy^2\right)dy \\
		&\stackrel{\mathclap{\alpha \ll 1}}{\approx}\; \; x - 2\alpha \int^x \left( vy + \frac{1}{2}u y^2 \right) dy\\
		&= \; \; x - \alpha v x^2 - \frac{\alpha}{3}ux^3.
	\end{aligned} 
\end{equation}
Plugging this into Eq.~\eqref{eq:rep_fix}, we obtain
\begin{equation}\label{eq:fix_freq}
	\begin{aligned}
		\frac{S(x)-S(0)}{S(1)-S(0)} &= \frac{S(x)}{S(1)} = \frac{x(1-\alpha vx - \frac{\alpha}{3}ux^2)}{1-\alpha v -\frac{\alpha}{3}u}\\
		&= x\left(1 - \alpha vx - \frac{\alpha}{3}ux^2 + \alpha v + \frac{\alpha}{3}u\right) + O(\alpha^2)\\
		&= x\left(1 + \alpha u \left(\frac{v}{u} (1-x) + \frac{1}{3} (1-x^2)\right)\right) + O(\alpha^2)\ .
	\end{aligned} 
\end{equation}
In the context of evolutionary game theory, this result is a re-derivation of the $1/3-$law \citep{nowak:Nature:2004} (generalized by \citet{lessard:JMB:2007}). It states that for an allele starting with one individual, it is more likely to become fixed in the population than under neutral dynamics if the deterministic fixed point is smaller than $1/3$. 
This can be seen by plugging in $u=a-b-c+d$ and $v=b-d$, where $a,b,c,d$ represent the payoffs of an evolutionary game.

\subsubsection*{Mean time to fixation}
The mean time to fixation is given by Eq.~\eqref{eq:ext_time} that was given as
\begin{equation}\label{eq:rep_ext}
	\begin{aligned}
		\mathbb{E}_x[\tau] &= \int_0^1 G(x,y) dy \\
		&= \int_0^x 2\frac{(S(1)-S(x))}{(S(1)-S(0))} \frac{(S(y)-S(0))}{\sigma^2(y)S'(y)} dy +  \int_x^1 2 \frac{(S(x)-S(0))}{(S(1)-S(0))}\frac{(S(1)-S(y))}{\sigma^2(y)S'(y)} dy\ ,
	\end{aligned}
\end{equation}
where $G(x,y)$ is Green's function and defined as (Eq.~\eqref{eq:Green})
\begin{equation}
	G(x,y) = \left\{ \begin{array}{ll} 2\frac{S(x)-S(0)}{S(1)-S(0)} (S(1)-S(y))m(y), & 0\leq x\leq y\leq 1, \\ 2 \frac{S(1)-S(x)}{S(1)-S(0)}(S(y)-S(0))m(y), & 0\leq y\leq x\leq 1, \end{array} \right. 
\end{equation}

Similar to the computation of the fixation probability, we will consider the case of small initial frequencies and weak selection, i.e. $\alpha,x\ll 1$. More precisely, we neglect terms of order $\alpha^2$ and $\alpha x^2$. We recall the approximation of the scale function in this case that we derived in Eq.~\eqref{eq:freq_scale} 
\begin{equation}
	S(x) \stackrel{\alpha \ll 1}{\approx} x - \alpha v x^2 - \frac{\alpha}{3}ux^3 \stackrel{x\ll 1}{\approx} x. 
\end{equation}
Employing these approximations, the first integral in Eq.~\eqref{eq:rep_ext} yields
\begin{equation}
	\begin{aligned}
	 &\int_0^x \left[ 2\frac{(S(1)-S(x))}{S(1)-S(0)} \frac{(S(y)-S(0))}{\sigma^2(y)S'(y)} \right] dy \\
	 &\quad = 2 \frac{S(1)-S(x)}{S(1)} \int_0^x \left[ \frac{y \left(1 -\alpha v y - \frac{\alpha}{3}u y^2\right)}{y (1-y) (1-2\alpha v y - \alpha u y^2)} \right] dy \\
	 &\quad \approx 2\left(1-\frac{x}{\left(1-\alpha v - \alpha \frac{u}{3}\right)}\right) \int_0^x \left[ \frac{\left(1-\alpha\left( vy + \frac{u}{3}y^2\right) \right) \left( 1+\alpha (2vy +uy^2)\right)}{(1-y)} \right] dy \\
	 &\quad \approx 2\left( 1-x\left( 1+\alpha\left( v+\frac{u}{3}\right)\right) \right) \int_0^x \left[ \frac{1-\alpha\left(vy +\frac{u}{3}y^2 -2vy -uy^2\right)}{(1-y)} \right] dy \\
	 &\quad \approx 2(1-x) \left(\int_0^x \frac{1}{1-y} dy + \underbrace{\int_0^x \frac{\alpha y \left(v+\frac{2}{3}y\right)}{1-y} dy}_{\in O(\alpha x^2)} \right) \\
	 &\quad \approx 2 (1-x) \ln((1-x)^{-1})\ .
	\end{aligned} 
\end{equation}
Approximating the second integral in a similar way we find 
\begin{equation}
	\begin{aligned}
	&\int_x^1 2 \frac{S(x)}{S(1)} \frac{(S(1)-S(y))}{\sigma^2(y)S'(y)} dy \\
	&\quad = 2 S(x) \int_x^1 \left[ \left(1- \frac{S(y)}{S(1)}\right) \frac{1}{y(1-y)(1-2\alpha v y-\alpha u y^2)}   \right]  dy \\
	&\quad \approx 2 S(x) \int_x^1 \left[ \left(1-\frac{y(1- \alpha v y - \frac{\alpha}{3}uy^2)}{1 -\alpha\left(v+\frac{u}{3}\right)}\right) \frac{(1+\alpha(2 v y+ u y^2))}{y(1-y)}  \right] dy \\
	&\quad \approx 2S(x) \int_x^1 \left[ \left(1-y\left(1 - \alpha v y - \frac{\alpha}{3}uy^2\right)\left(1+\alpha\left(v+\frac{u}{3}\right)\right)\right) \frac{(1+\alpha y (2 v + u y))}{y(1-y)} \right]dy\\
	&\quad \approx 2S(x) \int_x^1 \left[ \left(1-y - \alpha y \left( v (1-y) - \frac{u}{3}(1-y^2) \right) \right) \frac{(1+\alpha y (2 v + u y))}{y(1-y)} \right]dy\\
	&\quad \approx 2S(x) \int_x^1 \left[ \frac{1}{y} + \alpha (2v + uy) - \frac{\alpha\left(v(1-y)+\frac{u}{3}(1-y^2)\right) }{1-y} \right] dy \\
	&\quad = 2 S(x) \int_x^1 \left[\frac{1}{y} + \alpha\left( 2v + uy - v -\frac{u}{3}(1+y)\right)\right] dy \\
	&\quad \approx 2x \ln(x^{-1}) +2x\alpha \left(v(1-x) + \frac{u}{3}(1-x^2) - \frac{u}{3} (1-x)\right) \\
	&\quad \approx 2x \ln(x^{-1}) + 2x\alpha v\ .
	\end{aligned}
\end{equation}

Taking these two expressions together we re-derived the results already known in the literature \citep{altrock:NJP:2009, pfaffelhuber:TPB:2018}, i.e.
\begin{equation}
	\mathbb{E}_x[\tau] \approx 2(1-x)\ln((1-x)^{-1}) + 2x\ln(x^{-1}) + 2x \alpha v,
\end{equation}
for $\alpha$ and $x$ sufficiently small.

\end{appendix}

\end{document}